\documentclass[pre,twocolumn]{revtex4-1}

\usepackage{graphicx}
\usepackage{amsmath}
\usepackage{mathtools}
\usepackage{amssymb}

\usepackage{mathtools,cancel}
\usepackage{graphicx} 
\usepackage{booktabs} 
\usepackage{MnSymbol}

\usepackage{verbatim}   
\usepackage{color}      
\usepackage{subfigure}  
\usepackage{hyperref}   

\usepackage{mhchem}

\newcommand{\ba}{\begin{eqnarray}}
\newcommand{\ea}{\end{eqnarray}}
\newcommand{\be}{\begin{equation}}
\newcommand{\ee}{\end{equation}}

\begin{document}

\title{General theory of charge regulation within the Poisson-Boltzmann framework: 
study of a sticky-charged wall model} 

\author{Derek Frydel}
\affiliation{Department of Chemistry, Federico Santa Maria Technical University, Campus San Joaquin, Santiago, Chile}
\email[]{dfrydel@gmail.com}

\date{\today}

\begin{abstract}
This work introduces a sticky-charge wall model as a simple and intuitive representation of charge regulation. 
Implemented within the mean-field level of description, the model modifies the boundary conditions without 
affecting the underlying Poisson-Boltzmann (PB) equation of an electrolyte. Employing various modified PB 
equations, we are able to assess how various structural details of an electrolyte influence charge regulation.
\end{abstract}

\maketitle

\section{Introduction}

This work introduces a sticky-charged wall model as an intuitive
tool for describing charge regulation, a view that a surface
charge is not fixed but an outcome of two competing dynamic processes
at an interface: ion dissociation and ion binding. In consequence,
the effective surface charge is a function of the environment
in which it is embedded.  

The sticky potential is an idealization of a short-range surface
potential that arises from nonelectrostatic interactions, such as the
van der Waals interactions in the case of physisorption or the covalent
bindings in the case of chemisorption. It is formally obtained by
taking the width of the square-well potential to zero and its depth to
infinity \cite{Baxter68}.  By eliminating the range of the potential, the 
surface stickiness
is characterized by a single parameter; any microscopic details
are suppressed. Consequently, ions interact with a surface only
on direct contact. This leads to a simple proportionality between
the number of adsorbed ions and the contact density of an ionic
species.

Within the mean-field level of description, the sticky-charged
wall model modifies the boundary conditions without affecting
the underlying Poisson-Boltzmann (PB) equation. This transforms
the study of charge regulation to the obtaining of a solution
to a Poisson-Boltzmann equation with an alternative boundary 
conditions. By considering a number of different modified Poisson-
Boltzmann equations \cite{Frydel16}, we carry out a systematic study of how
different descriptions of an electrolyte influence adsorption.

The sticky-wall model, furthermore, can be modified to incorporate
a limited number of binding sites by renormalizing the
parameter of stickiness. This leads to the boundary conditions as
formulated within the Ninham-Parsegian model of charge regulation
based on chemical equilibrium and formulated in terms of
equilibrium constants \cite{Ninham71}.  

One of the goals of this work is to study the effects of the
solvation energy on the behavior of adsorption. One contribution
to the solvation energy is the dielectric constant of a solvent,
which screens the electric field and arises as the result of an orientational
polarization. Within the standard Poisson-Boltzmann
model, the orientational polarization is linear in the electric field,
which generally is accurate for low values of field or surface charge.
For large surface charges, nonlinear contributions start to modify
the dielectric constant in the vicinity of a surface, which, in turn,
modifies adsorption. The contributions of nonlinear polarization
to the structure of a double-layer had been investigated in Refs.
\cite{David07} and \cite{Frydel16}. The present paper extends this study into behavior of
adsorption.

In addition to the solvation effects, we study ion-specific effects
of polarizable ions on the behavior of adsorption. To this end, we
use the polarizable Poisson-Boltzmann equation \cite{Frydel11} 
and consider an electrolyte that is a mixture of polarizable and 
nonpolarizable ions.

This paper is organized as follows. In Sec. (\ref{sec:baxter}), we briefly overview
the Baxter sticky potential and review its derivation as the limiting
case of a square-well potential. In Sec. (\ref{sec:PB}), we consider the sticky-charged
wall model for the standard Poisson-Boltzmann equation.
In the same section, we consider modifications of the sticky surface
by limiting the number of sticky sites in order to make contact
with the Ninham-Parsegian model. In Sec. (\ref{sec:DPB}), we investigate
the contributions of the solvation energy, due to nonlinear polarization,
to the behavior of adsorption. To this end, we use the
dipolar Poisson-Boltzmann and the Langevin Poisson-Boltzmann
equations. In Sec. (\ref{sec:PPB}), we study the ion-specific effects of polarizable
ions and their effects on preferred adsorption. In this case, we use
the polarizable Poisson-Boltzmann equation. Finally, in Sec. (\ref{sec:con}), we
close the work with conclusions.

\section{Baxter sticky potential}
\label{sec:baxter}

The first implementation of a sticky potential is generally credited to Baxter who studied  
thermodynamic properties of sticky hard-spheres \cite{Baxter68}.  The adhesive 
interactions between spheres can give rise to clusters \cite{Bietry18} and, as clusters 
grow and combine, eventually lead to phase transition.  Since then, the sticky hard-sphere 
model has been explored by numerous groups, and today, it is 
considered as a good representation of colloidal gels \cite{Fuchs00,Piazza07,Royall18}.  
In electrostatics, the Baxter potential was first used to capture surface adsorption within 
the restricted primitive model \cite{Carnie81}.  In the series of following papers, 
\cite{Blum86,Blum89,Cornu89,Blum90,Blum90b} and 
\cite{McQuarrie95,McQuarrie96a,McQuarrie96b}, various theoretical aspects and 
variations of the sticky model were explored, most of the work carried out within the 
liquid-state theories.

In this work, we consider a sticky wall model within the framework
of the Poisson-Boltzmann equation. The absorbing surface is
considered to be planar and uniformly sticky with the sticky potential
denoted as $u_s(x)$. Due to its infinitesimal range, the Boltzmann
factor of the sticky potential is given by
\be
e^{-\beta u_s(x)} = 1 + l_s\delta(x), 
\label{eq:us}
\ee
where the length $l_s$ determines the strength of the surface stickiness.
(The sticky wall is assumed to be at $x = 0$.) The above result is most
conveniently obtained from the limit of the square-well potential, 
\be
  \beta u_{\rm well}(x) = \left\{ 
  \begin{array}{l l}
     -\varepsilon, & \quad \text{for ~~$0<x<a$}\\
     0, & \quad \text{for ~~$x>a$}.  
  \label{eq:U}
  \end{array} 
  \right.
\ee
By defining the finite range delta function, 
\be
  \Delta_a(x) = \left\{ 
  \begin{array}{l l}
     \frac{1}{a}, & \quad \text{for ~~$0<x<a$}\\
     0, & \quad \text{for ~~$x<0$ or $x>a$}, 
  \label{eq:U}
  \end{array} 
  \right.
  \ee
such that 
$
\lim_{a\to 0}\Delta_a(x)=\delta(x), 
$
the square-well potential is written as $\beta u_{\rm well}(x) = a\epsilon \, \Delta_a(x)$, 
and its Boltzmann factor becomes 
\be
e^{-\beta u_{\rm well}(x)} = 1 + a(e^{\varepsilon}-1)\Delta_a(x).  
\label{eq:lim_eu}
\ee
Taking the limit $a\to 0$ and $\varepsilon\to \infty$, such that $ae^{\varepsilon}=l_s$, 
the above result recovers Eq. (\ref{eq:us}), 
\be
\lim_{a\to 0}e^{-\beta u_{\rm well}(x)} = 1 + ae^{\varepsilon}\delta(x).  
\ee

Based on Eq. (\ref{eq:us}) we deduce that the total density $\rho(x)$ is split into the 
density of mobile particles, $\rho(x>0)$ (assuming the system is confined to a half-space 
$x>0$), and the density of adsorbed particles, $l_s\rho(0^+)\delta(x)$, which tells us that 
the surface density of adsorbed particles is $l_s\rho(0^+)$.  Any condition that alters the 
contact density $\rho(0^+)$ and the near-field region of a double-layer should, therefore, 
modify the adsorption behavior.

The link between the near-field region and the adsorption
behavior is later investigated in this work. As previous studies of
modified Poisson-Boltzmann equations do not indicate significant
far-field effects under condition in which the mean-field theory is
acceptable, there is an indication that near a surface these effects are
not negligible \cite{Frydel16}. This raises the possibility for the adsorption to be
dependent on the microscopic details of an electrolyte.

\section{The standard PB equation}
\label{sec:PB}

We start by considering a sticky-charged wall model 
within the standard Poisson-Boltzmann equation.  Here ions are represented
as structureless point-charges, and solvent as a background medium with a dielectric 
constant.  A sticky-charged wall is placed at $x=0$ where it confines an electrolyte 
to the region $x>0$.  The mean-field density of an ionic species $i$ is given by 
\be
\rho_i(x) = c_i e^{-\beta q_i \psi(x)} e^{-\beta u_i^{s}(x)},
\ee
where $\psi(x)$ is an electrostatic potential, $u_i^{s}(x)$ is the sticky interaction specific 
to a species $i$, and $q_i$ and  $c_i$ is the charge and the bulk concentration of a species 
$i$, respectively.  Since the Boltzmann factor of the sticky potential is 
\be
e^{-\beta u_i^s(x)} = 1 + l_i\delta(x), 
\ee
where $l_i$ is the parameter of stickiness for a species $i$, the density becomes 
\be
\rho_i(x) = c_i e^{-\beta q_i \psi(x)}\Big[1 + l_i \delta(x)\Big].  
\label{eq:rho_i}
\ee
Inserting this into the Poisson equation, 
\be
\epsilon \frac{d^2\psi(x)}{dx^2} = -\sum_{i=1}^K q_i \rho_i(x) - \sigma_c \delta(x), 
\ee
where $\sigma_c$ is the bare surface charge before adsorption, 
and $K$ is the total number of ionic species, we arrive at the Poisson-Boltzmann equation
\ba
\epsilon \frac{d^2\psi(x)}{dx^2} &=& -\sum_{i=1}^K q_i c_i e^{-\beta q_i\psi(x)} \nonumber\\
&-& \bigg[\sigma_c + \sum_{i=1}^K l_i q_i c_i e^{-\beta q_i\psi(0^+)} \bigg] \,\delta(x),
\label{eq:PB}
\ea
where $\psi(0^+)$ indicates the contact value of a potential from an electrolyte side.  
The sole difference between the above equation and the standard Poisson-Boltzmann 
equation is the modified surface charge in square parenthesis.

Alternatively, the effective surface charge can be incorporated into the boundary conditions. 
The expression of the boundary conditions is obtained by 
operating on Eq. (\ref{eq:PB}) with $\lim_{a\to 0}\int_{-a}^{a}dx$.  As the 
potential across a wall is continuous, $\psi(0^+)=\psi(0^-)$, this yields 
\be
\epsilon \frac{d\psi(x)}{dx}\bigg|_{x=0^+} = -\sigma_c - \sum_{i=1}^K q_i l_i c_i e^{-\beta q_i\psi(0^+)}.  
\label{eq:mbc1}
\ee
Within the region $x>0$, an electrolyte is governed by the standard Poisson-Boltzmann
equation, 
\be
\epsilon \frac{d^2\psi(x)}{dx^2} = -\sum_{i=1}^K q_i c_i e^{-\beta q_i\psi(x)}.  
\label{eq:pb1}
\ee
We note that $\psi(x<0)=\psi(0^+)$, which ensures continuity of an electrostatic potential
across a wall.

The resulting boundary conditions in Eq. (\ref{eq:mbc1}) suggests that the effective surface 
charge is 
\be
\sigma_{\rm eff} = \sigma_c + \sum_{i=1}^K q_i l_i c_i e^{-\beta q_i\psi(0^+)}, 
\label{eq:sigma_eff}
\ee
where the second term accounts for a surface charge due to adsorbed ions.  If 
adsorbed ions have the same charge as the bare surface charge, the effective surface 
charge will increase. For the opposite situation, it will be reduced.

In consequence, the effective surface charge is not fixed, or
determined a priori, as we do not know the value of a potential at
a contact with a wall, $\psi(0^+)$. It can only be determined by solving
Eq. (\ref{eq:pb1}). The mathematical coupling of the differential equation to
the boundary condition reflects the physical coupling between the
surface and an electrolyte, which are in contact and in dynamic
equilibrium. Any variation in an electrolyte will be reflected in the
effective surface charge. Such coupling is a feature and consequence
of charge regulation.

The boundary conditions in Eq. (\ref{eq:mbc1}) are also interesting from
mathematical point of view. It combines the values of a function $\psi(x)$
and the values of its derivative on the boundary, however, not in a
linear manner, which does not qualify it as the usual Robin boundary
condition but can be considered as a highly nonlinear generalization
of it.

The solution of the Poisson-Boltzmann equation for a fixed surface
charge is obtained by the iterative numerical procedure, starting
with an initial guess for an electrostatic potential and then correcting
that solution using exact internal relations that check for self-consistency.
The same method is used for the sticky wall model. The effective surface 
charge $\sigma_{\rm eff}$ is simply corrected after each iteration 
using the value of $\psi(0^+)$ from the previous iteration. For the case
of counterion adsorption, each consecutive correction reduces the
effective surface charge and, as a result, charge regulation accelerates
convergence and facilitates the numerics.

\subsection{alternative interpretation of a sticky potential}

In the same way as we define the effective surface charge $\sigma_{\rm eff}$ whose 
value depends on the number of adsorbed ions, we can define the effective stickiness, 
whose value changes with the number of adsorbed particles.  The microscopic origin of 
such renormalization is that adsorption occurs at specific discrete sites, where each 
site can bind with at most one ion.  Once occupied, the sites are deactivated and the 
averaged surface density of active sites becomes reduced.  This, in turn, leads to 
the reduction in a wall stickiness.  

If the total number of sites per unit area is $\gamma_i$ (under the assumption that sites
are ion specific) and none of the sites are occupied, the effective stickiness parameter is 
$l_{\rm eff}=l_i$.  On the other hand, if all the sites are occupied, $l_{\rm eff}=0$.
Within the same logic, half the sites occupied implies $l_{\rm eff}=l_a/2$.  This suggests 
the following expression for the effective parameter of stickiness 
\be
l_i^{\rm eff} = l_i\bigg( \frac{\gamma_i-l_i^{\rm eff} \rho_i(0^+)}{\gamma_i^{\rm}}\bigg), 
\label{eq:l_eff}
\ee
where $l_i^{\rm eff} \rho_i(0^+)$ is the surface density of adsorbed ions of 
the species $i$, and $\gamma_i-l_i^{\rm eff} \rho_i(0^+)$ is the number of unoccupied
sites.  
Solving the above equation for $l_i^{\rm eff}$ we get  
\be
l_i^{\rm eff}
= \frac{l_i\gamma_i}{\gamma_i+ l_ic_ie^{-\beta q_i \psi(0^+)}}, 
\label{eq:l_eff2}
\ee
and the new boundary conditions are 
\be
\epsilon \frac{d\psi(x)}{dx}\bigg|_{x=0^+} = -\sigma_c 
- \sum_{i=1}^K  \frac{q_i l_i c_i e^{-\beta q_i\psi(0^+)} }{1+ l_i c_i e^{-\beta q_i\psi(0^+)}/\gamma_i}.  
\label{eq:mbc2}
\ee
The surface is now characterized by the parameters $\sigma_c$, $\{l_i\}$, and $\{\gamma_i\}$.

Other versions of a sticky wall are possible.  For example, if the binding sites bind indiscriminately 
to all ions, the corresponding boundary conditions are 
\be
\epsilon \frac{d\psi(x)}{dx}\bigg|_{x=0^+} = -\sigma_c 
- \sum_{i=1}^K  \frac{q_i l_i c_i e^{-\beta q_i\psi(0^+)} }{1+ \sum_{i=1}^K  l_i c_i e^{-\beta q_i\psi(0^+)}/\gamma}.  
\label{eq:mbc3}
\ee
In this case, the surface is characterized by the parameters $\sigma_c$, $\{l_i\}$, and $\gamma$.

We refer to the sticky boundary conditions in Eq. (\ref{eq:mbc1}), Eq. (\ref{eq:mbc2}), and 
Eq. (\ref{eq:mbc3}) as the boundary condition $A$, $B$, and $C$.  In the subsequent work, we 
consider mainly the boundary condition of the type $A$.

The boundary conditions $B$ and $C$ are the same as those obtained in 
the Ninham-Parsegian model of charge regulation \cite{Ninham71,David16,David18}, derived 
based on the formalism of chemical equilibrium, where adsorption 
is determined by the dissociation constant 
$K_i^{\rm d}=\gamma_i/l_i$.

The expressions in Eq. (\ref{eq:mbc2}) and (\ref{eq:mbc3})  for the boundary conditions
"B" and "C" have similar structure as the modified Poisson-Boltzmann equation in 
Ref. \cite{David97}, which accounts for the excluded volume effects using a local 
approximation; that is, it applies expressions of a homogeneous system to a heterogeneous 
situation \cite{Frydel12}.  In consequence, a local approximation fails to produce the usual 
density oscillations seen in exact systems and leads to unphysical local
density saturation.

The saturation of Eq. (\ref{eq:mbc2}) and Eq. (\ref{eq:mbc3}) implies that the number of 
adsorbed particles is bound from above by the value of $\gamma_i$.  The saturation in 
this case, however, occurs only within the plane of a sticky surface, thus, the local 
approximation is applied to a single point at $x=0$ and not the entire region.   
Consequently, there is no saturation within the density profile 
itself.  This is seen if we consider that the density of ions at a surface is represented by 
a delta function, $l_i^{\rm eff}\rho_i(0^+)\delta(x)$, which clearly precludes any possibility 
of local saturation. 

One can raise the question about the role of distribution of biding
sites on the behavior of adsorption. For example, such sites could
be distributed randomly or on a regular lattice. To explore such possibilities
would require a number of careful simulations for different
distributions of the binding sites. As the goal of this work is to focus
on the boundary conditions "A", and the cases "B" and "C" are mentioned
to make contact with other models of charge regulation, we
do not undertake such a detailed study.

\section{Effect of the solvation energy on adsorption}
\label{sec:DPB}

In this part of the paper, we look into how the change in the solvation energy 
due to nonlinear polarization of a solvent modifies  
adsorption.  The expected trend is that the more favorable solvation energy 
facilitates ion dissolution, thereby, reduces adsorption.  
Large dielectric constant, which implies increased screening of electrostatic
interactions, is associated with better solvation energy.  In consequence, if the
nonlinear polarization near a surface leads to increased dielectric
constant, the adsorption will be reduced.  If it leads to reduced dielectric constant,
the adsorption will be enhanced.  

In the standard PB equation, the polarization density of a solvent is linear in electrostatic 
field, $P(x)=-\chi_e\psi'(x)$, where $\chi_e$ is the electric susceptibility, and the resulting 
dielectric constant is $\epsilon=\epsilon_0(1+\chi_e)$.  
Nonlinear contributions to $P(x)$ could arise in strong 
electrostatic field as a result of orientational saturation, when molecular
dipole is aligned with a field and fails to respond to its further increase.  
Furthermore, if a solvent is compressible, nonlinearities could arise due to 
accumulation of the solvent dipoles near a surface.

\subsection{The Dipolar PB equation}

We start with the dipolar Poisson-Boltzmann equation (DPB).  
The DPB model was first conceived as a more realistic representation of 
a dipolar solvent \cite{David07}.  It represents solvent particles as a gas of 
point-dipoles.  The nonlinearities in this model are primarily the result of inhomogeneous 
distribution of a solvent.

To derive the DPB model, we consider the polarization density, $P(x)=\rho_d(x)p(x)$, 
where $\rho_d(x)$ is the density of a dipolar species and $p(x)$ is the local dipole 
moment of a single solvent molecule.  The polarization density contributes to the 
charge density as $-P'(x)$, and the Poisson equation is 
\be
\epsilon_0 \frac{d^2\psi(x)}{dx^2} = -\sum_{i=1}^K q_i \rho_i(x) + \frac{d P(x)}{dx}, 
\ee
and, for a sticky-charged wall case, the boundary conditions are 
\be
\epsilon_0 \frac{d\psi(x)}{dx}\bigg|_{x=0^+} = -\sigma_c + P(0^+) - \sum_{i=1}^K l_i q_i \rho_i(0^+).  
\ee
In the above equations, $\epsilon_0$ is the dielectric constant of a vacuum.  For the linear 
polarization, $P(x)=-\chi_e\psi'(x)$, we recover the standard PB equation with the dielectric 
constant $\epsilon=\epsilon_0(1+\chi_e)$.  Note that the 
effective surface charge, in addition to adsorbed ions, includes the polarization 
surface density, $P(0^+)$.

The DPB equation is obtained by substituting for $\rho(x)$ and $P(x)$ their respective
mean-field expressions given by $\rho_i(x)=c_ie^{-\beta q_i\psi(x)}$ and \cite{David07,Frydel16}
\be
P(x) = \bigg(\frac{c_d \sinh \big(p_0\beta \psi'(x)\big)}{p_0\beta\psi'(x)}\bigg)
\bigg( p_0 L\big(p_0\beta\psi'(x)\big)\bigg), 
\label{eq:P}
\ee
where $c_d$ is the bulk concentration of a dipolar species, $p_0$ is the permanent dipole 
moment, and $L(x)= \coth(x) - x^{-1}$ is the Langevin function.  The terms in parenthesis 
correspond to $\rho_d(x)$ and $p(x)$, respectively.

\begin{figure}[h] 
 \begin{center}
 \begin{tabular}{rrrr}
  \includegraphics[height=0.19\textwidth,width=0.23\textwidth]{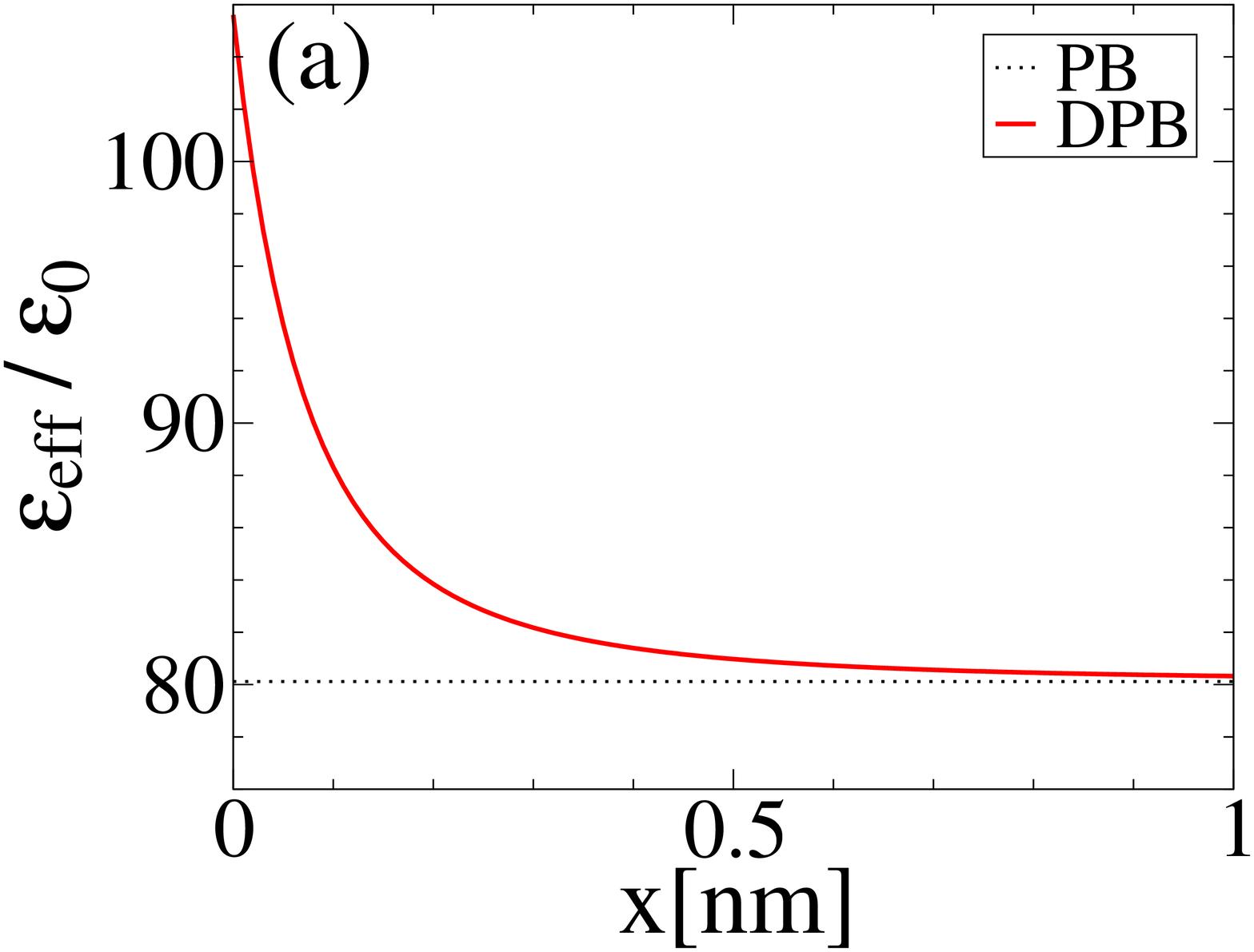}&
  \includegraphics[height=0.19\textwidth,width=0.23\textwidth]{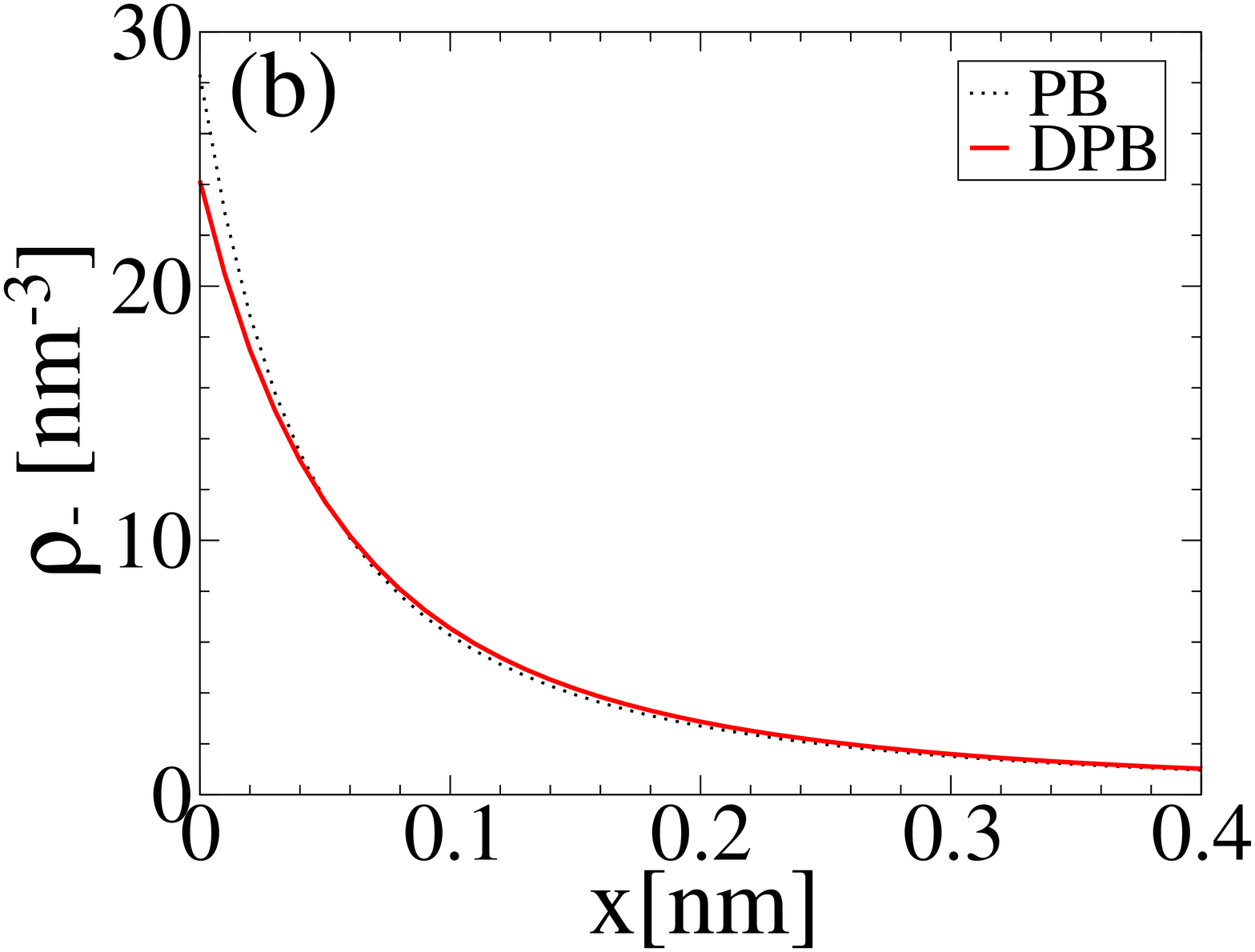}
 \end{tabular}
 \end{center}
\caption{(a) the effective dielectric constant, $\epsilon_0 - \frac{P(x)}{\psi'(x)}$, and (b)
the counterion density as a function of a distance from a charged (non-sticky) wall.  
The solvent parameters for the DPB model are $c_d=55~{\rm M}$ and $p_0=4.78~{\rm D}$ 
such that far away from a wall, where $P(x)\sim - \psi'(x)$, the effective dielectric constant
recovers the value of water, $\epsilon = 1+\beta c_d p_0^2/3 \approx 80\epsilon_0$.  
The remaining parameters are $c_s=0.1~{\rm M}$, $\sigma_c=0.4~{\rm Cm^{-2}}$. }
\label{fig:eps_DPB} 
\end{figure}

Before considering a sticky wall case, we briefly look into a non-sticky wall model to see 
how the nonlinear polarization of the DPB model modifies the system with reference to the 
standard PB model.  In Fig. (\ref{fig:eps_DPB}) we plot the effective dielectric constant,
defined as
\be
\epsilon = \epsilon_0 - \frac{P(x)}{\psi'(x)}, 
\ee
and the density of counterions for a non-sticky wall for a symmetric $1:1$ electrolyte.  
The parameters $p_0$ and $c_d$ correspond roughly to those of water, such that the 
linear regime, $P(x) \approx -\frac{\beta c_d p_0^2}{3}\psi'(x)$, recovers the value of 
water, $\epsilon/\epsilon_0 \approx 80$.  The nonlinear contributions, strongest near 
a wall, enhance the dielectric constant and the electronic screening as solvent 
accumulates near a wall.  In consequence, the counterion density, in the region around a 
wall is reduced. Based on this result, we expect the counterion adsorption to become 
reduced as a consequence of more favorable solvation energy in the vicinity of a wall.  
\graphicspath{{figures/}}

\begin{figure}[h] 
 \begin{center}
 \begin{tabular}{rrrr}
  \includegraphics[height=0.19\textwidth,width=0.23\textwidth]{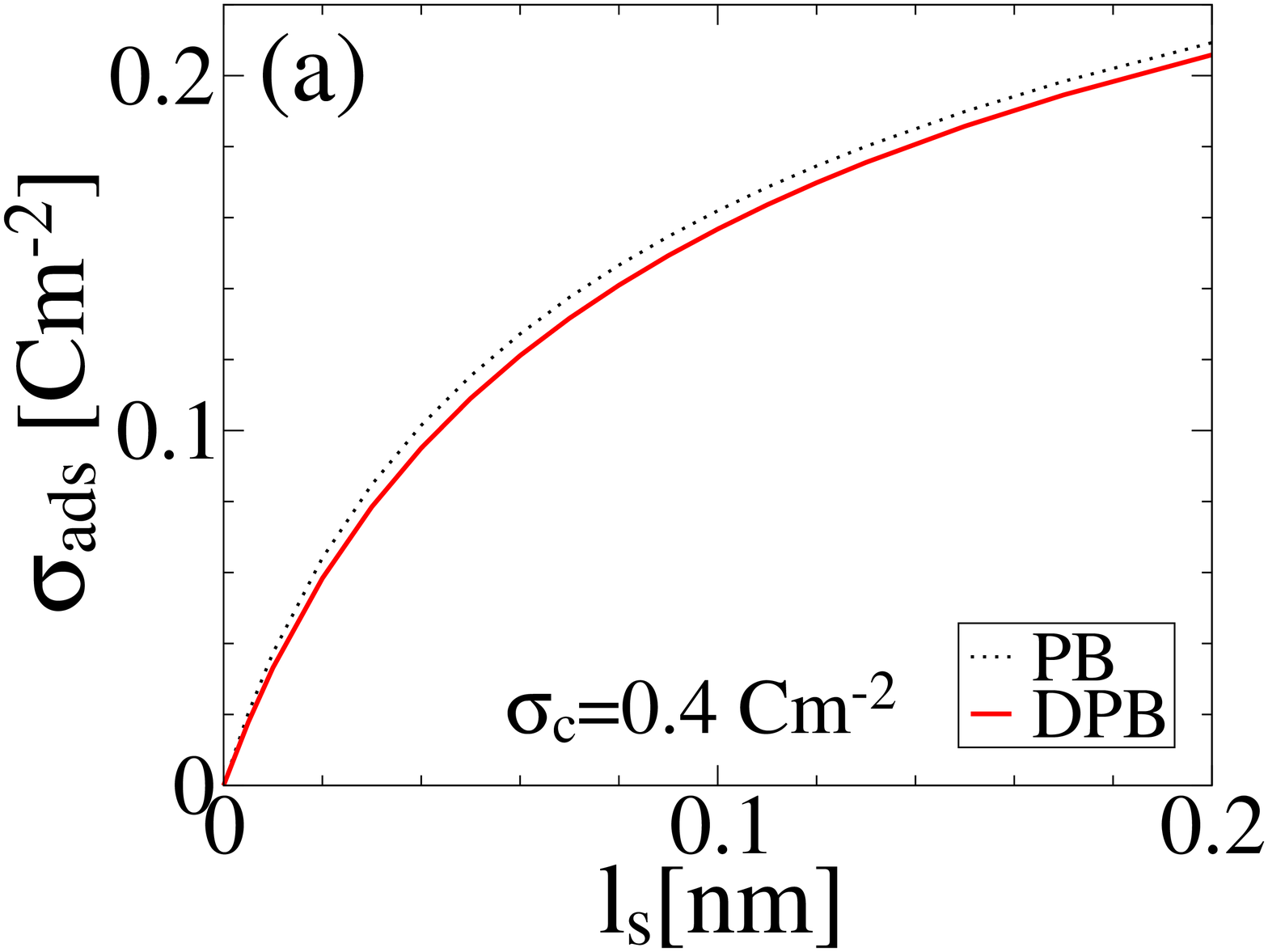}&
  \includegraphics[height=0.19\textwidth,width=0.23\textwidth]{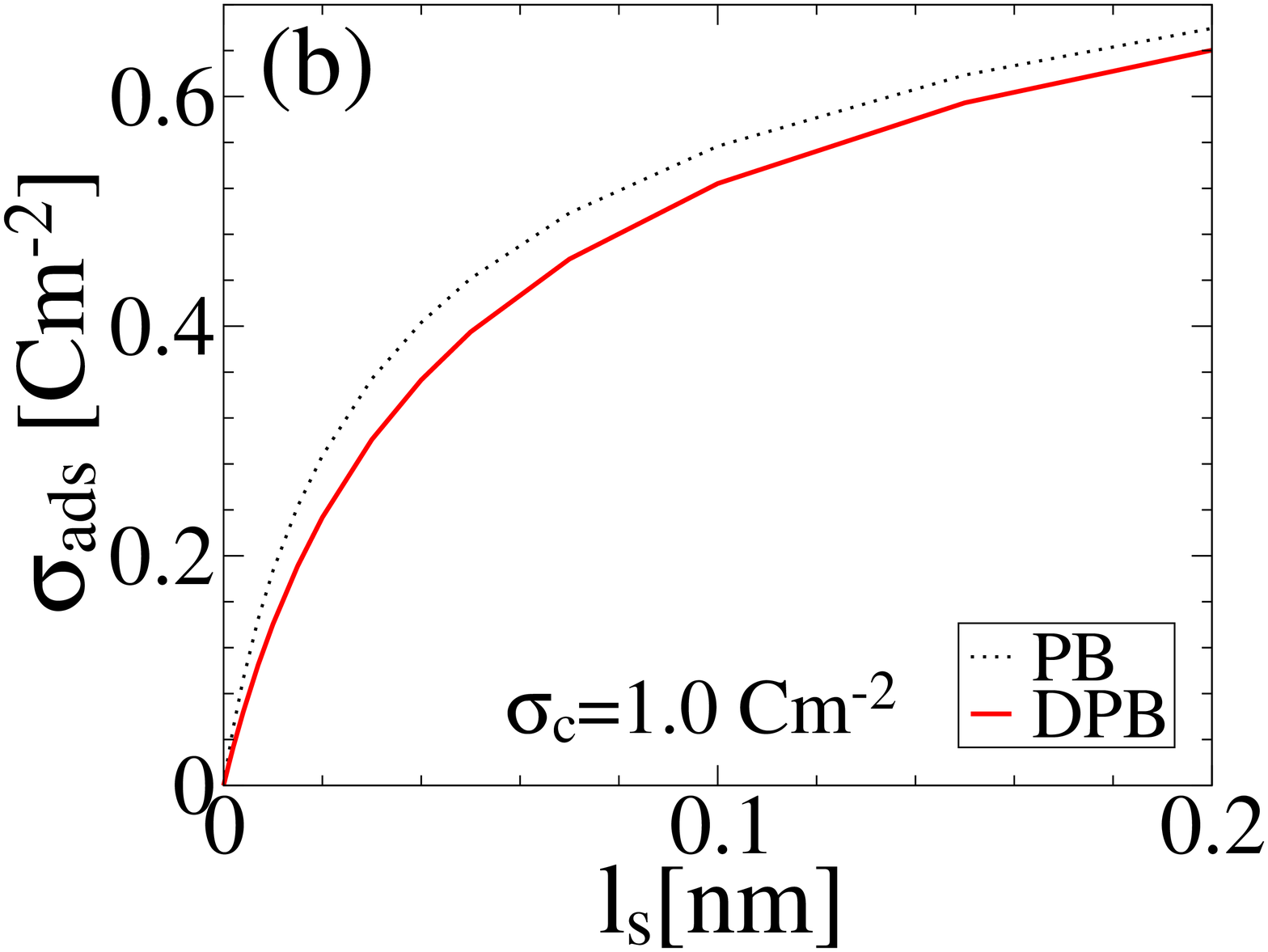}
 \end{tabular}
 \end{center}
\caption{The surface charge density of adsorbed ions for the PB and the DPB model,
for the surface charge density (a) $\sigma_c=0.4~{\rm Cm^{-2}}$ and (b) $\sigma_c=1.0~{\rm Cm^{-2}}$.
The remaining system parameters are as those in Fig. (\ref{fig:eps_DPB}).  }
\label{fig:sig_ads_DPB}
\end{figure}

To verify this assertion, in Fig. (\ref{fig:sig_ads_DPB}) we plot the surface charge 
density of adsorbed ions, $\sigma_{\rm ads}$, as a function of the wall stickiness, $l_s$, 
for two different values of a surface charge density, $\sigma_c$.  The parameter of 
stickiness are the same for each species, $l_s=l_+=l_-$.  As anticipated, the adsorption 
is reduced (in relation to the PB model with the same parameters).  
Since nonlinear effects become stronger for large values of electric field, the reduction
in adsorption is more pronounced as $\sigma_c$ increases.  
\graphicspath{{figures/}}

\subsection{The Langevin PB equation}
\label{sec:LPB}

In the DPB model, the dipolar solvent is represented as compressible.  
This leads to a rather large and unphysical accumulation of a dipolar 
species near a wall.  A more realistic representation of water solvent 
should assume incompressibility.  This can be done by setting the dipolar 
density as  
\be
\rho_d(x) = c_d, 
\ee
leading to  
\be
P(x) = c_d p_0 L\big(p_0\beta\psi'(x)\big), 
\label{eq:P2}
\ee
and we refer to this model as the Langevin PB equation (LPB) \cite{Frydel16}.  

\begin{figure}[h] 
 \begin{center}
 \begin{tabular}{rrrr}
  \includegraphics[height=0.19\textwidth,width=0.23\textwidth]{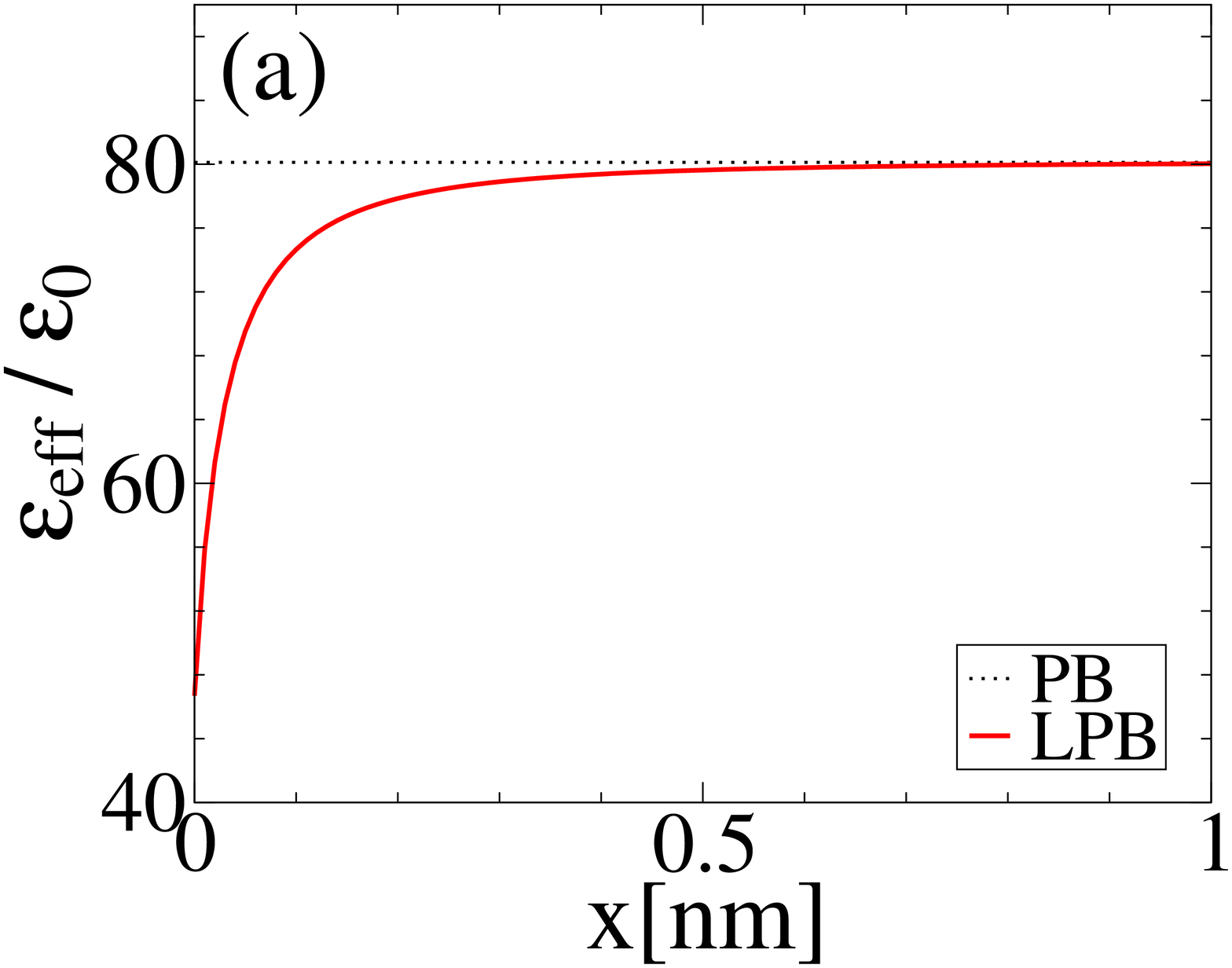}&
  \includegraphics[height=0.19\textwidth,width=0.23\textwidth]{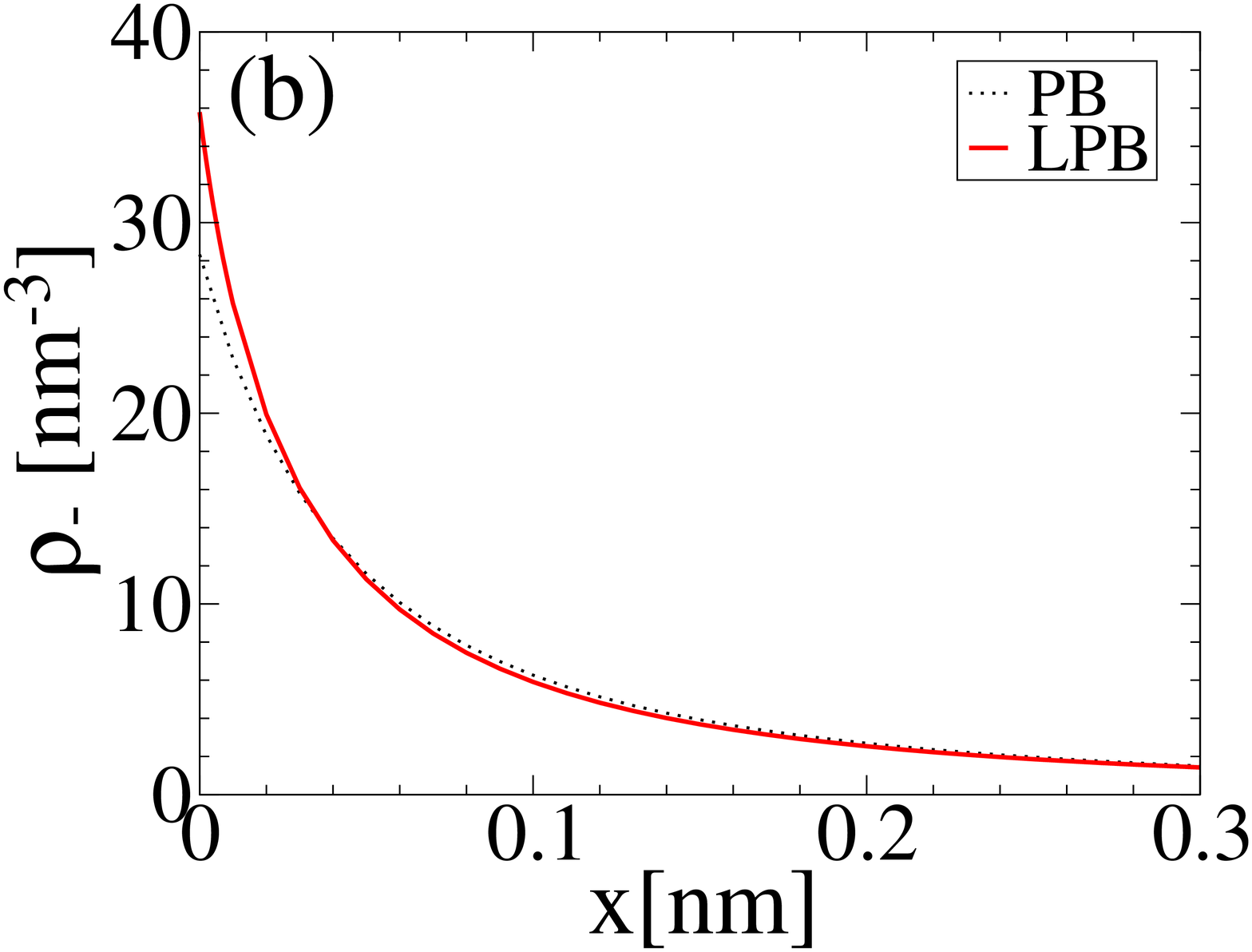}
 \end{tabular}
 \end{center}
\caption{(a) The effective dielectric constant and (b) the counterion density 
as a function of a distance from a charged (non-sticky) wall, for the LPB model
and the parameters as those in Fig. (\ref{fig:eps_DPB}).  }
\label{fig:eps_LPB} 
\end{figure}

Because an incompressible solvent cannot accumulate at a surface, we expect, 
in contrast to the DPB model, a reduced electrostatic screening in consequence to the 
saturation effect of the Langevin function in Eq. (\ref{eq:P2}).  This is demonstrated
by Fig. (\ref{fig:eps_LPB}) (a) where the dielectric constant is reduced in the region 
near a wall.  This, in turn, leads to increased density shown in Fig. (\ref{fig:eps_LPB}) (b).   
For a sticky wall model this implies increased adsorption.  
\graphicspath{{figures/}}

\begin{figure}[h] 
 \begin{center}
 \begin{tabular}{rrrr}
  \includegraphics[height=0.19\textwidth,width=0.23\textwidth]{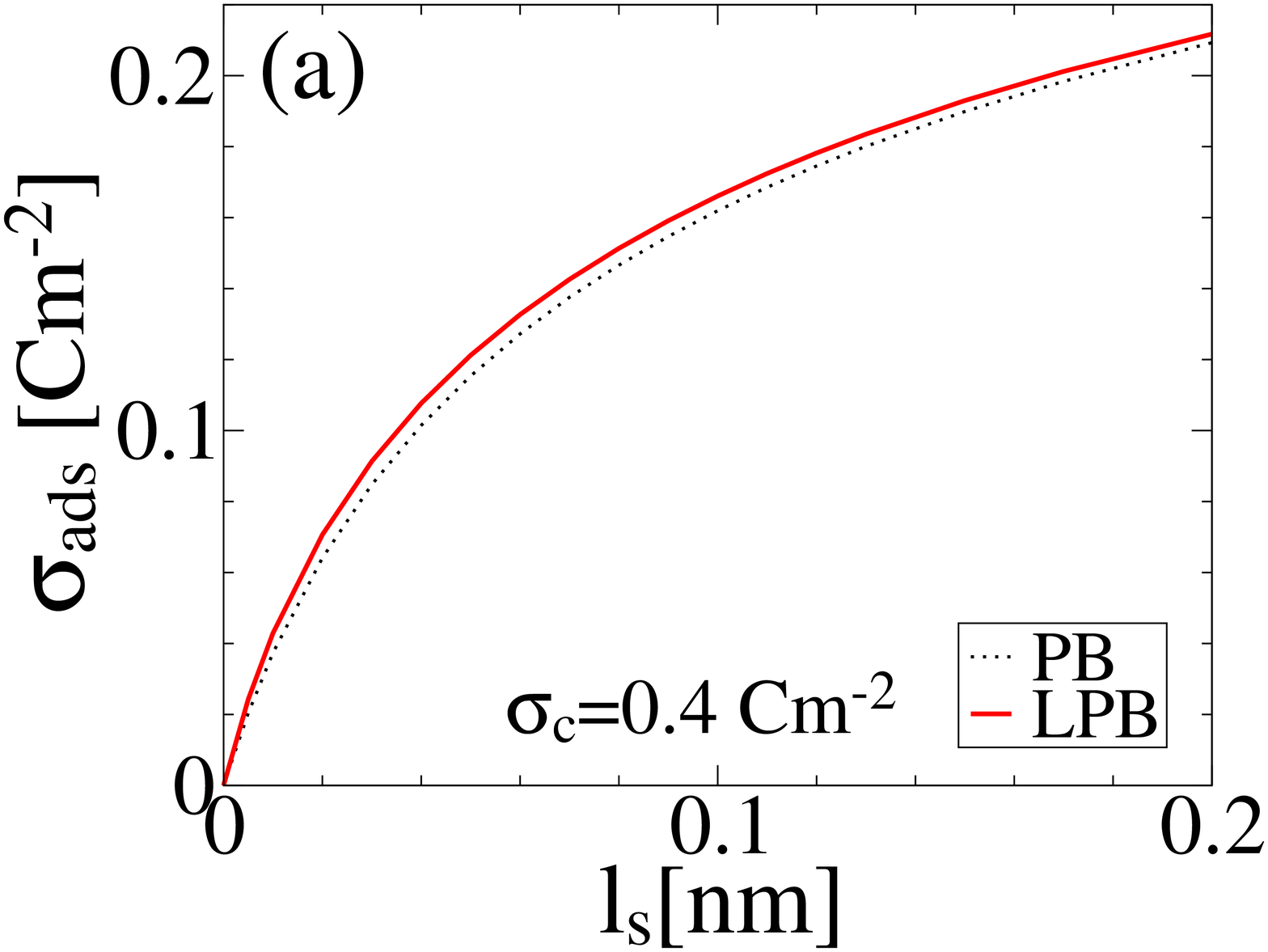}&
  \includegraphics[height=0.19\textwidth,width=0.23\textwidth]{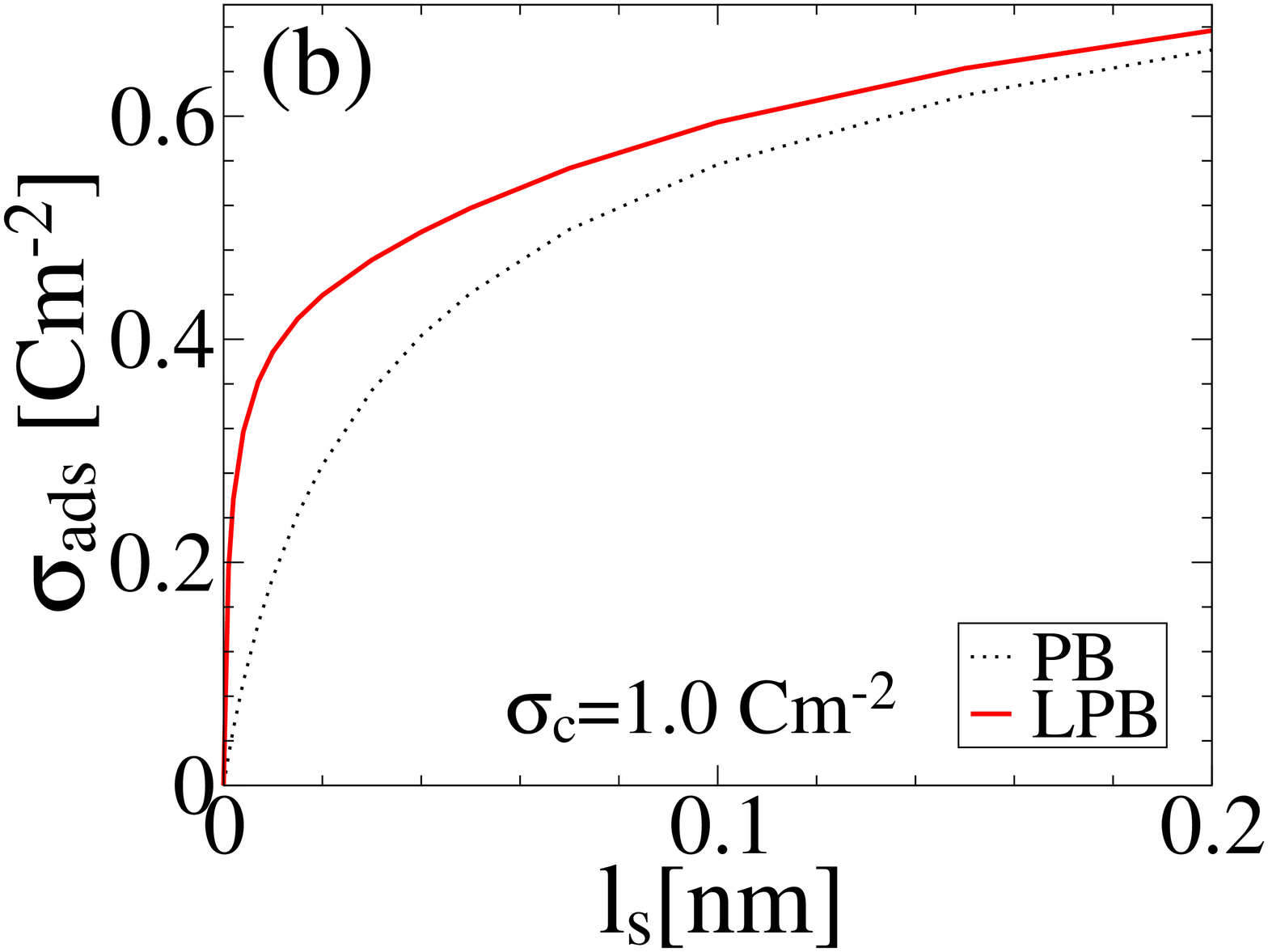}&
 \end{tabular}
 \end{center}
\caption{The surface charge density of adsorbed ions as a function of $l_s$ for the 
LPB model, for the surface charge density (a) $\sigma_c=0.4{Cm^{-2}}$ and (b)
$\sigma_c=1.0{\rm Cm^{-2}}$.  The remaining system parameters are the same as 
those in Fig. (\ref{fig:eps_LPB}).  }
\label{fig:sig_ads_LPB} 
\end{figure}

When we next look into the adsorption behavior of a sticky-wall model, see 
Fig. (\ref{fig:sig_ads_LPB}), we encounter
\graphicspath{{figures/}}
enhanced adsorption in relation to the 
standard PB model with the same sticky parameters.  The nonlinear polarization 
effects become larger with increasing $\sigma_c$.

We note that the reduced effective dielectric constant for the LPB model 
shown in Fig. (\ref{fig:eps_LPB}), and the accompanying increase of electrostatic
interactions, implies increased role of correlations absent in the mean-field
description \cite{Yan02}.  The question of correlations for a system that incorporates 
explicit solvent is of course complex.  The frequently used random-phase 
approximation in electrostatics \cite{Frydel16b}, often presented in the field-theoretical 
formalism as the Gaussian approximation \cite{Frydel15}, is for the present situation 
not very trustworthy, given that it does not stand a test of 
comparison even for as simple systems as the Gaussian core \cite{Frydel17} 
or the penetrable sphere model \cite{Frydel18}.  If correlations were present, 
it is expected that they should enhance adsorption as a result of counterion 
ordering in the longitudinal plane \cite{Yan02}, thus, the predictions of the present 
mean-field model can be regarded as the lower bound of what one would see in
a more accurate approximation.

\section{The polarizable PB equation and ion-specific effects}
\label{sec:PPB}

In this section we consider 
the polarizable Poisson-Boltzmann equation (PPB) \cite{Frydel11,Frydel16}, 
which represents ions as polarizable point charges, and whose mean-field density is 
given by 
\be
\rho_i(x) = c_ie^{-\beta q_i\psi(x)}e^{\beta\alpha_i\psi'^2(x)/2}, 
\label{eq:rho_ppb}
\ee
Note that 
any finite polarizability $\alpha_i$ increases the concentration of ions in the presence of
an external field.  This indicates that ions with larger polarizability are more likely 
to be found near a charged surface where electrostatic field is larger.  By the same 
token, polarizable ions are more likely to be adsorbed.

The PPB equation is written as 
\be
\epsilon \frac{d^2\psi(x)}{dx^2} = -\sum_{i=1}^K q_i \rho_i(x)  
- \frac{d}{dx}\bigg[ \psi'(x)\sum_{i=1}^K\alpha_i \rho_i(x)\bigg],  
\ee
where the second term on the right hand side comes from the polarization density
due to polarizable ions, 
\be
P(x) = -\psi'(x) \sum_{i=1}^K\alpha_i \rho_i(x).  
\ee
Then the boundary conditions for the sticky wall model are given by 
\be
\epsilon \frac{d\psi(x)}{dx}\bigg|_{x=0^+}  
= -\sigma_c - \psi'(0^+) \sum_{i=1}^K\alpha_i \rho_i(0^+) - \sum_{i=1}^K l_i q_i \rho_i(0^+).  
\ee

Because adsorbed ions are polarizable and an external field can induce dipole moments, 
the boundary conditions need to account for a surface density of dipoles.  Surface 
dipoles are accounted by a discontinuous electrostatic potential across a surface, 
and the discontinuity that arises is given by  
\ba
\psi(0^+) - \psi(0^-) &=& -\psi'(0^+)\frac{1}{\epsilon}\sum_{i=1}^K l_i  \alpha_i \rho_i(0^+), 
\label{eq:dphi}
\ea
where it lowers the potential 
in the region outside an electrolyte, $\psi(x<0)$, but has no effect on the region inside an 
electrolyte, consequently, it does not effect adsorption.  A jump in the electrostatic 
potential could be relevant for fluid interfaces with trapped colloids, where it 
could affect the structure of a double-layer around colloids \cite{Frydel07,Frydel11b}.

\begin{figure}[h] 
 \begin{center}
 \begin{tabular}{rrrr}
  \includegraphics[height=0.19\textwidth,width=0.23\textwidth]{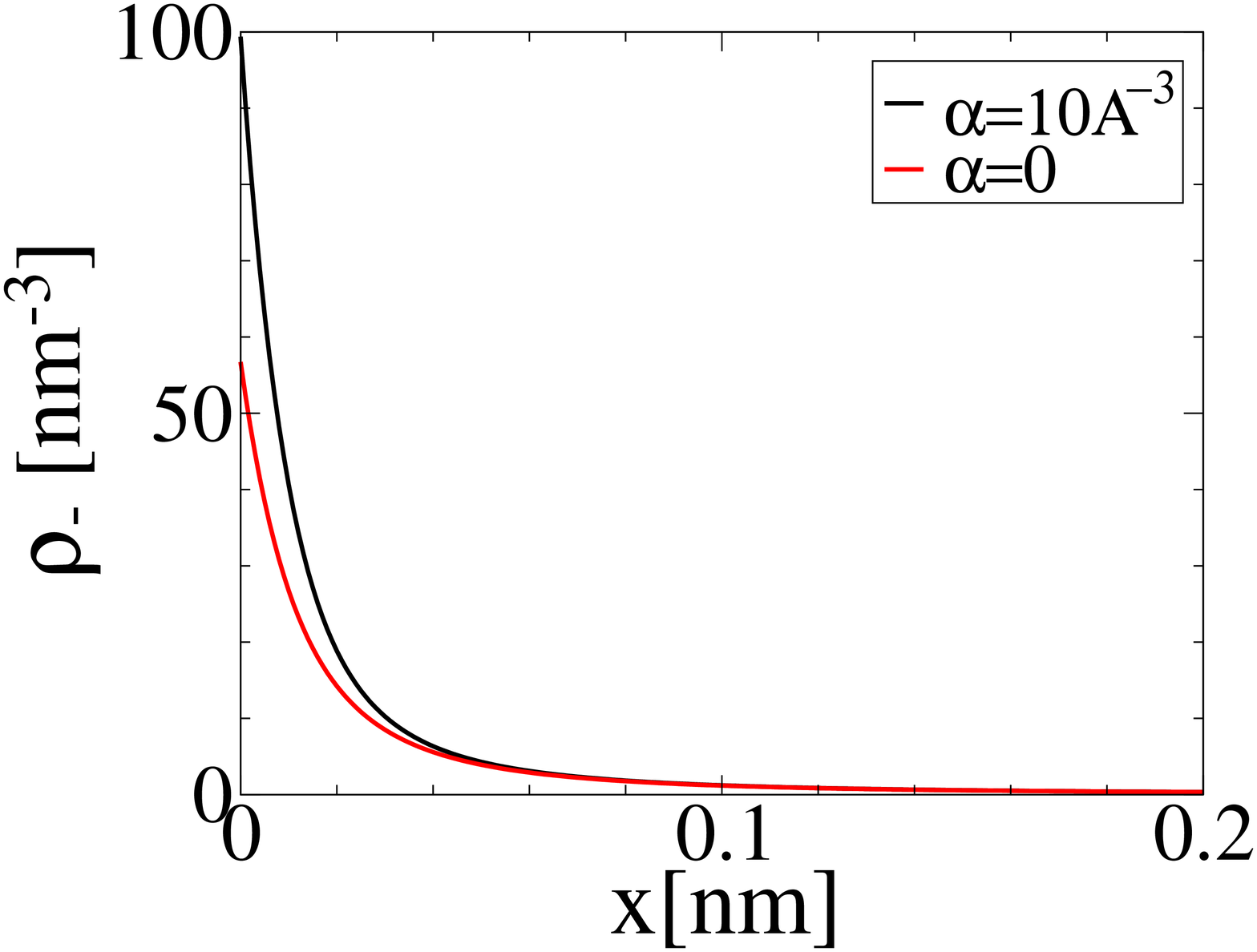}&
 \end{tabular}
 \end{center}
\caption{The counterion density as a function of a distance from a charged (non-sticky) 
for a $50:50$ mixture of polarizable ($\alpha/4\pi\epsilon_0=10{\rm \AA^3}$,
which roughly corresponds to that of an iodide ion) and non-polarizable ions.  The 
remaining parameters are $c_s=0.1{\rm M}$, $\sigma_c=0.4{\rm Cm^{-2}}$, 
and $\epsilon/\epsilon_0=10$, which corresponds to $\lambda_B=5.76{\rm nm}$
(where $\lambda_B=\beta e^2/4\pi\epsilon$ is the Bjerrum length). }
\label{fig:rho_PPB2} 
\end{figure}

To investigate ion specific effects, we consider a system in which half of 
counterions are polarizable, and the other half is non-polarizable.  
We start with a non-sticky wall to determine the distribution of different
counterions within a double-layer around a wall. These distributions are shown 
in Fig. (\ref{fig:rho_PPB2}) for a solvent with a low dielectric constant, 
$\epsilon/\epsilon_0=10$, in order to emphasize the ion-specific effects.  
For solvents with higher dielectric constant, such as water, the ion-specific 
effects due to ion polarizability are small.  This implies that these effects 
are relevant for systems with low dielectric constant, such as
ionic fluids, and less so for aqueous solutions.  
\graphicspath{{figures/}}

\begin{figure}[h] 
 \begin{center}
 \begin{tabular}{rrrr}
  \includegraphics[height=0.19\textwidth,width=0.23\textwidth]{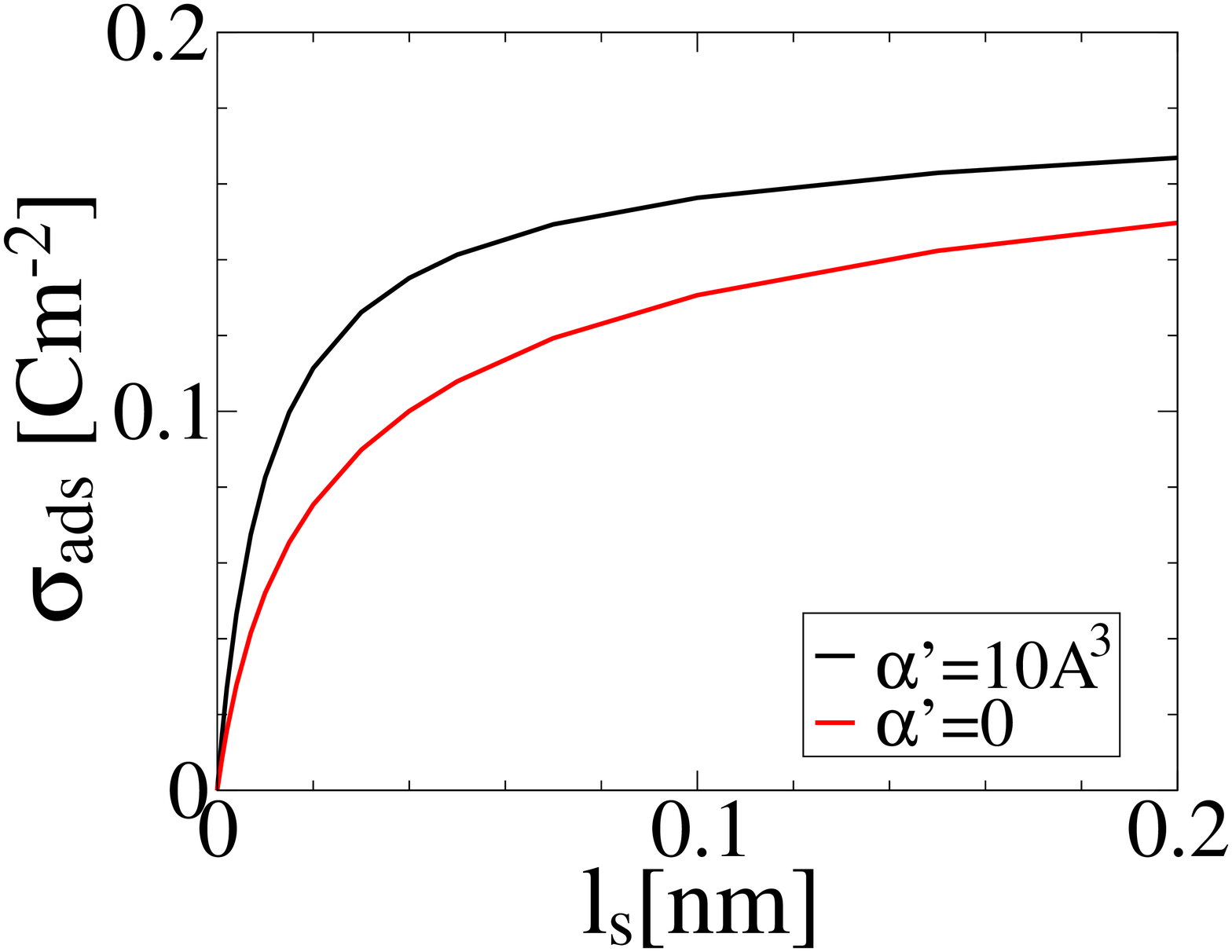}&
 \end{tabular}
 \end{center}
\caption{Surface charge density of adsorbed ions as a function of $l_s$ for a mixture of
polarizable and non-polarizable ions.  The remaining parameters are as in 
Fig. (\ref{fig:rho_PPB2}).  The results indicate that polarizable ions are 
more likely to be adsorbed.  }
\label{fig:sig_ads_PPB2} 
\end{figure}

Next, we consider a sticky-wall model for the same system parameters.  
Fig. (\ref{fig:sig_ads_PPB2}) plots the surface density of adsorbed polarizable and 
non-polarizable counterions.  As expected, based on the results in Fig. (\ref{fig:rho_PPB2}), 
polarizable ions are more likely to be adsorbed than non-polarizable ones.  This can 
be traced to the fact that the density of polarizable ions, see Eq. (\ref{eq:rho_ppb}), 
have larger concentration for any non-zero electric field.  Physically this means that
the induction of a dipole moment lowers the energy.  
\graphicspath{{figures/}}

\section{Conclusion}
\label{sec:con}

In conclusion, we have considered the sticky-charged wall
model as a simple and intuitive tool to incorporate charge regulation.
The model is sufficiently flexible that it can be modified to
take into account a limited number of binding sites and allows us to
arrive by a different route at the Ninham-Parsegian model of charge
regulation \cite{Ninham71}. The dissociation constant in that model is inversely
proportional to the stickiness parameter in the present sticky wall
model.

In the later part of this work, we study various electrostatic
effects that may arise as a result of a more detailed microscopic
description of an electrolyte. We separate these contributions into
those due to the solvation energy and those due to the structure of
dissolved ions. In the former case, we consider nonlinear contributions
of a solvent due to compressibility and orientational saturation
in strong fields. As the saturation makes solvation near a wall unfavorable,
adsorption is enhanced as a consequence. For the case of
ion structure, we consider polarizability and its effect on adsorption.
The observed effect is that polarizable ions are more susceptible to
adsorption.

The effects described above are significant under extreme conditions,
that is, at large surface charges, where polarization saturation
is significant, and/or at a low dielectric constant of a solvent,
where polarizability effects become more pronounced. Within
the standard conditions and the weak-coupling limit, the microscopic
details produce only small variations. Consequently, this
study is particularly relevant for systems like ionic liquids where the
absence of a dielectric solvent makes electrostatic screening weak, in
which case variations in microscopic structure give rise to disparate
behaviors.

\begin{acknowledgments}
D.F. would like to thank David Andelman for introducing the topic of charge regulation.  
D.F. would also like to acknowledge discussions with Yan Levin 
on some aspects of boundary conditions.  
\end{acknowledgments}



\end{document}